\documentclass[conference,a4paper]{IEEEtran}
\usepackage{amsmath,graphicx,amsfonts}
\usepackage{xcolor}
\usepackage{amssymb}
\usepackage{subfigure}
\usepackage{cite}

\IEEEoverridecommandlockouts
\begin{document}
%
\title{Extreme Image Coding via Multiscale Autoencoders with Generative Adversarial Optimization}

\author{Chao Huang, Haojie Liu, Tong Chen, Qiu Shen, and Zhan Ma\IEEEauthorrefmark{1}\thanks{Corresponding Author: Z. Ma.}\\Vision Lab, Nanjing University}



%


\maketitle

\begin{abstract}
	We propose a MultiScale AutoEncoder (MSAE) based extreme image coding/compression framework to offer visually pleasing reconstruction at a very low bitrate. Our method leverages the ``priors'' at different resolution scale  to improve the compression efficiency, and also employs the generative adversarial network (GAN) with multiscale discriminators to perform the end-to-end trainable rate-distortion optimization. We compare the perceptual quality of our reconstructions with traditional compression algorithms using High-Efficiency Video Coding (HEVC) based Intra Profile and JPEG2000 on the public {\it Cityscapes}, {\it ADE20K} and {\it Kodak} datasets, demonstrating the significant subjective quality improvement. However, objective measurements, such as PSNR, SSIM, etc, are often deteriorated by applying the generative adversarial optimization.
	
\end{abstract}
%



%
\IEEEpeerreviewmaketitle

\section{Introduction}
\label{sec:intro}

{Images that capture vivid scenes and events are stored and shared extensively every day. Thus image compression plays a vital role to ensure the efficient storage and sharing at the entire Internet scale. Traditional image compression methods such as JPEG, JPEG2000, HEVC based BPG, as well as recent deep neural network (DNN) based image compression methods~\cite{toderici2017full,rippel2017real,balle2018variational,Liu2018DeepCoder} have presented significant advances in compression efficiency. Typically, these DNN-based schemes exhibit better visual quality than the traditional methodologies, at the same bit rate~\cite{balle2016end}. However, both of them fail to represent images efficiently with pleasant reconstruction quality at very low bitrates (e.g., targeting for $< 0.05$ bits per pixel (bpp))~\cite{agustsson2018generative}}.




\begin{figure*}[t]
	\centering
	\subfigure[]{\includegraphics[scale=0.16]{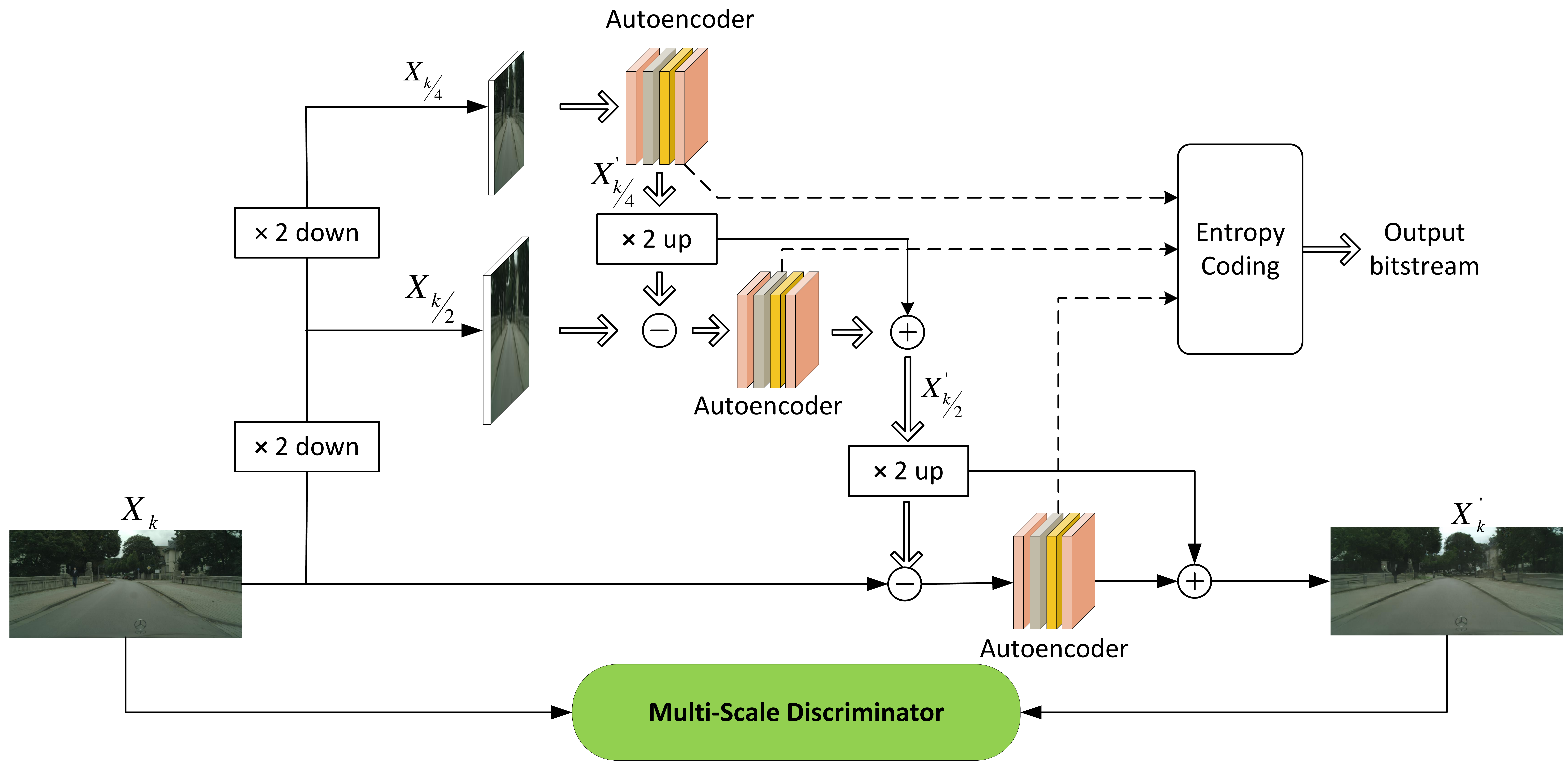} \label{fig:MSAE}}
	\subfigure[]{\includegraphics[scale=0.22]{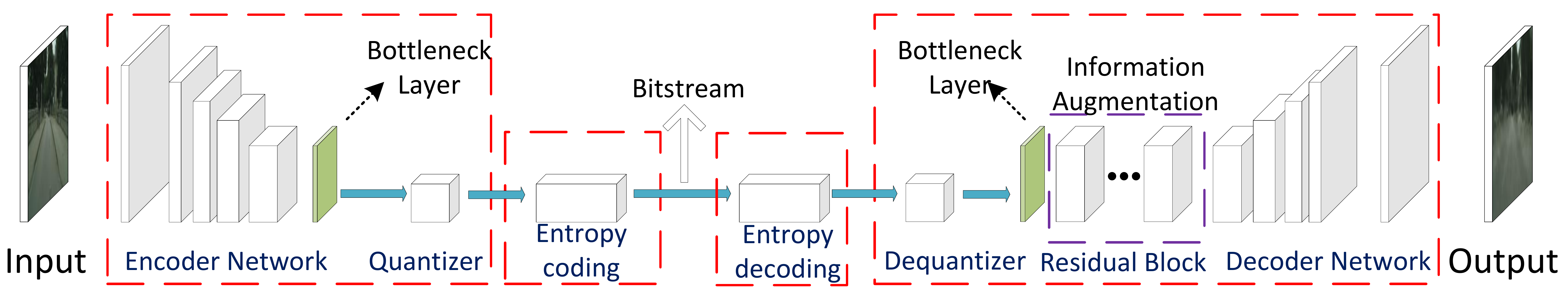} \label{sfig:AE}}
	\caption{Our extreme image compression framework via Multi-Scale AutoEncoder (MSAE) with GAN optimization. (a) overall structure, (b) autoencoder.
    {The multi-scale distriminator in (a) contains three identical discriminators that are patch-based fully convolutional networks~\cite{long2015fully}.}
		The encoder network contains 1 convolutional layer with stride 1 and 4 convolutional layers with stride 2; all the residual blocks in information augmentation have the same convolutional kernel size 3 and stride 1; the decoder network is a mirror version of the encoder, which contains 4 transposed convolutional layers with stride 2 and 1 convolutional layer with stride 1. Entropy encoding and decoding denote the arithmetic encoding and decoding.} 
	
\end{figure*}

{This is mainly due to the reason that visual sensitive information (i.e., perceptual significance) can not be well preserved using conventional quality optimization criteria, such as peak signal-to-noise ratio (PSNR) and multiscale structural similarity (MS-SSIM)~\cite{wang2003multiscale}, at such extreme compression scenario. Recent explorations have shown that adversarial loss could be a tentative solution to capture global semantic information and local texture, yielding appealing reconstructions~\cite{goodfellow2014generative,agustsson2018generative}. Thus, Agustsson {\it et al.}~\cite{agustsson2018generative} developed a GAN-based extreme image compression framework with bitrate below 0.1 bpp, resulting in the noticeable subjective quality improvement compared with the JPEG2000~\cite{taubman2012jpeg2000} and BPG~\cite{bpg}. However, it had limitations by adopting a purely GAN-based structure. First, it was difficult to ensure the generalization of GAN to capture a variety of distributions of different datasets. In the meantime, GAN sometimes would introduce unexpected textures because of the failure of discriminator~\cite{wang2018recovering}. }

{In this work, we propose a MultiScale AutoEncoder (MSAE) based extreme image compression structure where we employ a multiscale network shown in Fig.\ref{fig:MSAE} to generate spatial scalable bitstreams. To the best of our knowledge, most learning based compression methods~\cite{toderici2017full,rippel2017real,balle2018variational,Liu2018DeepCoder}, generate a single layer bitstream at its native spatial resolution, without utilizing mutual information from other spatial scales. Different from Scalable Auto-encoder~\cite{JiaScalableAE} that iteratively codes the pixel-level errors at the same resolution,  ``priors'' at different spatial resolution scale that well capture the local textures, are embedded as reference to help the coarse-to-fine reconstruction and compression in our MSAE framework.  Generative Adversarial loss~\cite{wang2018high} is applied in different scales for end-to-end trainable rate-distortion optimization, so as to optimize the reconstruction quality subjectively by maintaining the global semantic structure for visual significance, at a very low bit rate budget. We have our method tested on {\it Cityscapes}, {\it ADE20K} and {\tt Kodak} datasets, yielding significant perceptual quality margins over the existing JPEG2000 and BPG.
}

	
	
	\section{MultiScale AutoEncoder with Generative Adversarial Optimization}
	\label{sec:model}
	Fig.\ref{fig:MSAE} presents the extreme image compression framework of MSAE with generative adversarial optimization.  Let $X_k$ be the original image ($k$ is the size of the input). We downscale the $X_k$ to obtain two more inputs $X_{{k}/{s}}$ and $X_{{k}/{(s*s)}}$. $s$ denotes the downscaling factor, which is set by 2 in this paper. Let $\mathbb{A}_i$ be the autoencoder network at scale $i$ ($i\in[k,{k}/{2}, {k}/{4}]$), and $\mathbb{U}$ denotes the upscaling operator. We then define the overall MSAE framework by
	\begin{align}
	X_{{k}/{4}}^{'} &=\mathbb{A}_{{k}/{4}}(X_{{k}/{4}}), \label{eq:MASE_lscale}\\
	X_{{k}/{2}}^{'} &= \mathbb{U}(X_{{k}/{4}}^{'}) + \mathbb{A}_{{k}/{2}}(X_{{k}/{2}}-\mathbb{U}(X_{{k}/{4}}^{'})), \label{eq:MASE_mscale}\\
	X_{k}^{'} &= \mathbb{U}(X_{{k}/{2}}^{'}) + \mathbb{A}_{k}(X_{k}-\mathbb{U}(X_{{k}/{2}}^{'})). \label{eq:MASE}
	\end{align}
	
	Our proposed MSAE framework in \eqref{eq:MASE_lscale}, \eqref{eq:MASE_mscale}, and \eqref{eq:MASE} has presented a coarse-to-fine reconstruction step by step. At the lowest scale ${k}/{4}$, the autoencoder $\mathbb{A}_{{k}/{4}}$ only takes $X_{{k}/{4}}$  as an input to derive the reconstructed image $X_{{k}/{4}}^{'}$, yielding the {\it coarsest} representation of original $X_{k}$. Then $X_{{k}/{4}}^{'}$, as the prior,	 is upscaled and aggregated with residuals at each scale to derive the final $X_{k}^{'}$. Low resolution reconstructions are referred as ``priors'' to improve the overall rate-distortion performance.
	In addition, conditional GAN~\cite{mirza2014conditional} is integrated into our MSAE system to do end-to-end training for visually appealing reconstruction, by enabling the multiscale discriminators for each input high-resolution images.
	
	\subsection{AutoEncoder}
	\label{sec:AE}

	\begin{figure*}[ht]
		\centering
		\includegraphics[scale=0.185]{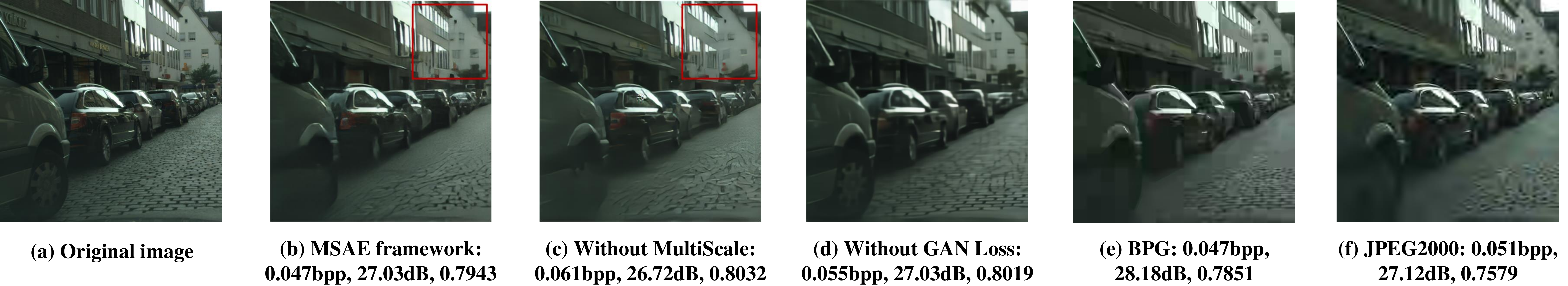}
		\caption{{Visual comparison on different architectures and loss functions, evaluated on a real-world image from Cityscapes dataset (values below each image are bitrate, PSNR and SSIM). In (c), we replace MSAE model with a single scale model. (d) MSAE with MS-SSIM loss optimization with color degradation. In (e) and (f), traditional codec frameworks make the reconstructed images have undesired blur and artifacts. Our complete model in (b) is able to produce better prediction. Compared with the result in (c), MSAE model can preserve local textures better (in red box).}}
		\label{sfig:comparison}
	\end{figure*}
	
	%

	The same autoencoder architecture is used in our MSAE framework at each scale. Except at the scale $k/4$ where the downscaled image $X_{k/4}$ serves as the input, residuals between upscaled priors and inputs (i.e., $X_{{k}/{2}} - \mathbb{U}(X_{{k}/{4}}^{'})$ and $X_k - \mathbb{U}(X_{{k}/{2}}^{'})$)  at the same resolution, are fed into the autoencoder for compression. Using residuals, instead of default textures, generally boost the coding efficiency at the same bitrate budget due to better energy compaction and redundancy exploit.
	
	Such autoencoder, shown in Fig.\ref{sfig:AE}, includes an encoder ${E}$ to encode the input $X$ to a set of feature maps (fMaps) $\omega$. Then the $\omega$ is passed to the quantizer $Q$ and will be quantized to a compressed representation $\hat{\omega}=Q(E(X))$.  Specifically, the encoder $E$ first compresses the input with size of $W \times H \times C$ to  feature maps with dimensions at $\frac{W}{16}\times\frac{H}{16}\times 480$. Usually, $W$ is for image width, $H$ is the height and $C$ is the number of color channels (e.g., $C=3$ for RGB color space). The fMaps are then projected down to $\frac{W}{16}\times\frac{H}{16}\times C_{neck}$ at bottleneck layer prior to being quantized for $\hat{\omega}$. Note that $C_{neck}$ varies at different scale.
	
	The decoder, denoted by the generator $G$, tries to reconstruct the image $X^{'}={G({\hat{\omega}})}$ from the compressed representation $\hat{\omega}$. Within the decoder, {\it information augmentation module} with nine residual blocks~\cite{he2016deep} is aggregated to retrieve more information from the data to improve the reconstruction. Decoded fMaps will go through a mirror network of $E$ to obtain final reconstruction with dimensions at the same dimension, i.e., $W \times H \times C$, as the input image.

	
	
	Note that the autoencoder is optimized using PSNR or MS-SSIM in default, often resulting in compression artifacts such as blocking, blurring and contouring effects at a low bitrate. To address this problem, we adopt adversarial loss~\cite{goodfellow2014generative} in training to reconstruct image ${X^{'}}$ with visually pleasant quality.
	


	
	\subsection{End-to-End Rate-Distortion Optimization}
	\label{sec:advtraining}
	
	
	We adopt adversarial training in end-to-end optimization framework for extreme compression.
	This is mainly due to the reason that adversarial loss can address the blurring and contouring problems at a low bitrate level~\cite{goodfellow2014generative}. In the proposed framework, the decoder or generator $G$ is conditioned on the compressed representations and there is no necessity to add random noise for generator~\cite{mirza2014conditional}. For discriminator $D$, we use the multiscale architecture following~\cite{wang2018high}, which measures the divergence between real image and fake image generated by $G$ both globally and locally. Here we introduce a loss function that is closer to the perceptual similarity instead of relying on pixel-wise distortion~\cite{ledig2017photo}, i.e.,
		\begin{align}
		L_{f} &= \nonumber\\
		&\frac{\lambda}{W_{m,n}H_{m,n}}\sum\limits_{x=1}^{W_{m,n}}\sum\limits_{y=1}^{H_{m,n}}\left(\phi_{m,n}(Y_k)_{x,y}-
		\phi_{m,n}(Y'_k)_{x,y}\right)^{2},
		\end{align}
	with $Y_k = D(X_{k})$ and  $Y'_k = D(X'_{k})$. $\phi_{m,n}$ represents  the feature map generated by the $n$-th convolution (with stride 2) of the $m$-th scale for the multiscale discriminator. $W_{m,n}$ and $H_{m,n}$ are the dimensional size of the respective feature maps. For the coefficient $\lambda$, we set it to 10.
	
	The regular GAN~\cite{goodfellow2014generative} hypothesizes the discriminator as a classifier with the sigmoid cross entropy loss function, which may lead to  gradient vanishing problem.
	In this paper, we use objective measures $f(y)=(y-1)^2$ and $g(y)=y^2$ developed for Least-Squares GAN~\cite{mao2017least}, where $f$ and $g$ denote the scalar functions. It results in the generator loss as,
		\begin{equation}
		L_{G} = \mathop {\min }\limits_{G} f\left(D\left(G(\hat{\omega}_{k})+\mathbb{U}(G(\hat{\omega}_{{k}/{2}})+\mathbb{U}(G(\hat{\omega}_{{k}/{4}}))\right)\right),
		\end{equation}
	and the discriminator loss as:
		\begin{equation}
		L_{D}=\mathop {\min }\limits_{D}\left( f\left(D(X_{k})\right) + g\left(D(X_{k}^{'})\right)\right).
		\end{equation}

	
	In order to backpropagate through the non-differentiable quantizer $Q$, we model the entropy rate following the~\cite{balle2018variational} at bottleneck layer. 
	We simply add uniform noise to ensure differentiability during training and replace it with {\rm ROUND}($\cdot$) in inference. The entropy of $\hat{\omega}_{i}$ is evaluated using:
		\begin{align}
		H(\hat{\omega}) = -\sum\limits_{j}\log_{2}(p_{\hat{\omega}_{j}\mid\psi^{(j)}}(\hat{\omega}_{j}\mid\psi^{(j)})),
		\end{align}
	where $\psi^{(j)}$ represents parameters of each univariate distribution $p_{\hat{\omega}_{j}}$.
	To balance the quality of the reconstruction and the bitrate, An entropy rate term need to be added to the training loss for optimal rate-distortion efficiency, i.e.,
		\begin{align}
		L_{\rm RD} &= \nonumber\\
		&\mathop {\min }\limits_{d,H}\sum\limits_{i\in[k,\frac{k}{s},\frac{k}{s*s}]}\left(  L_{G} +  \alpha_{i} d(X_{i},X_{i}^{'}) +
		L_{f} + \beta_{i}H(\hat{\omega}_{i})\right).
		\end{align}
	As we can see, the rate-distortion trade-off is adjusted by setting the variations of $\alpha_{i}$ and $\beta_{i}$.
	Distortion, i.e., $d(X_{i},X_{i}^{'})$, is measured by the PSNR in this study, and the entropy of compressed representation, i.e., $H(\hat{\omega}_{i})$, is used to approximate the encoding bitrate~\cite{balle2018variational}. Such compound loss $L_{\rm RD}$ is applied in a end-to-end trainable framework to achieve the optimal rate-distortion performance.

	\section{Experimental Studies}
	\label{sec:experiments}

	{\bf Datasets:}
	We use two public accessible datasets for training: Cityscapes~\cite{cordts2016cityscapes} and ADE20K~\cite{zhou2017scene}. Cityscapes dataset contains $3475$ images, each of them has the dimension of $2048\times1024\times3$ in RGB color space. During the training, we randomly select 2400 images for training and the rest for validation. These images are downscaled to $1024\times512\times3$ in our experiments to avoid GPU memory overflow in training. For the ADE20K dataset, we choose $4927$ images. It is then segmented randomly to a training set and a validation set, with sizes from $256\times256\times3$ to $1024\times1024\times3$. For simplicity, we rescale all of them to $512\times512\times3$ for training and validation.

	
	
	{\bf Parameters:}
	We set $\beta_{k}=100$ and $\beta_{{k}/{4}} = \beta_{{k}/{2}}=1$. Meanwhile, $\alpha_{k}=1$, $\alpha_{{k}/{4}} = \alpha_{{k}/{2}}=100$ accordingly. The number of channels of the bottleneck layer varies between different scales. For scale ${k}/{4}$ and ${k}/{2}$, we set the $C_{neck}=1$, while at scale $k$, $C_{neck}=4$. {This setting aims to provide sufficient ``prior" information while consuming less bit overhead.} Additionally, we use a learning rate of $2 \times 10^{-4}$ and the Adam optimizer for end-to-end learning.
	
	\begin{figure}[ht]
		\centering
		\includegraphics[scale=0.14]{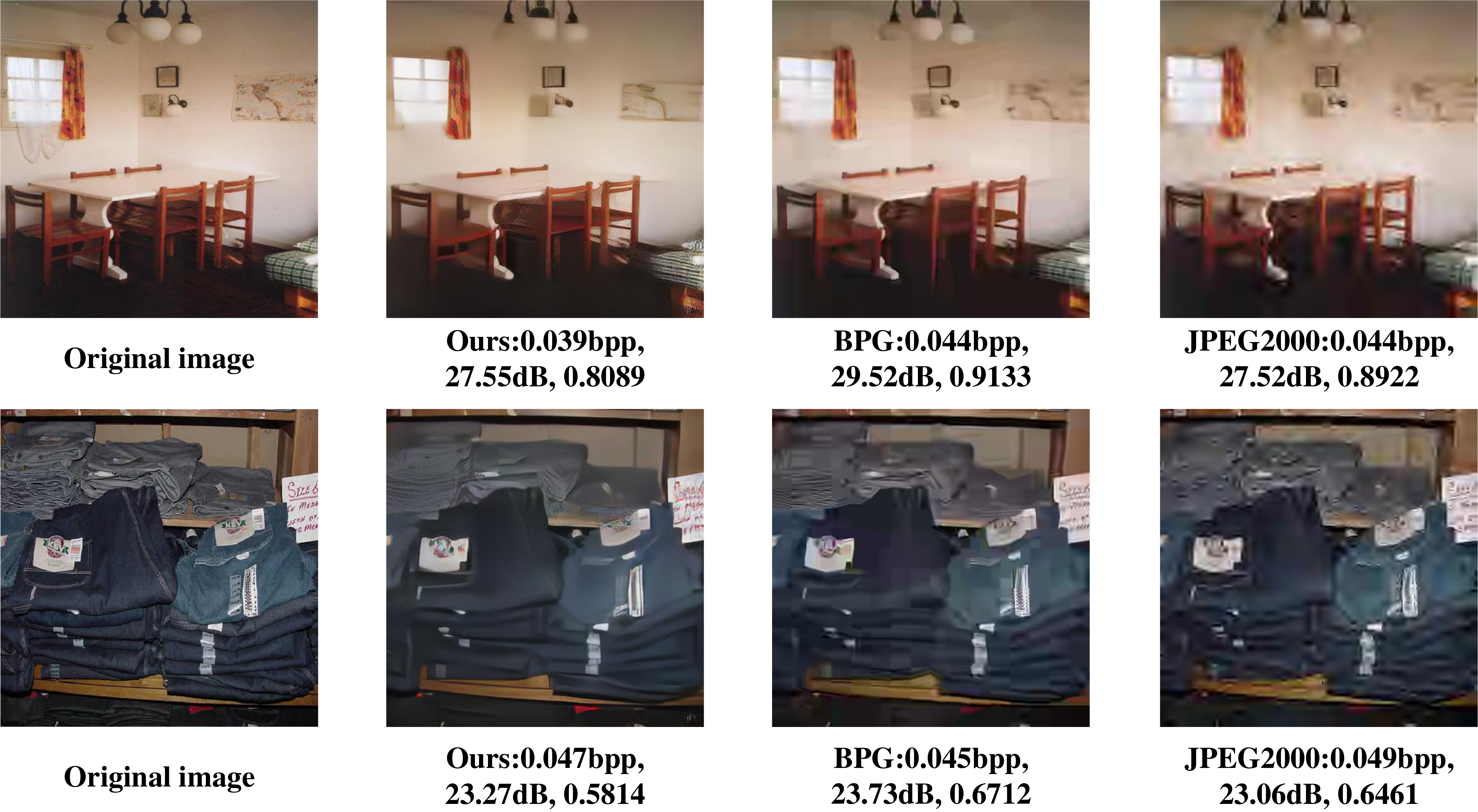}
		\caption{Illustration of performance comparison for our proposed extreme image compression method versus BPG, JPEG2000 on ADE20K dataset with objective PSNR, SSIM and subjective snapshots.}
		\label{sfig:ADE20K}
	\end{figure}
	
		\begin{figure}[ht]
		\centering
		\subfigure{
			\begin{minipage}[t]{0.29\linewidth}
				\centering
				\includegraphics[width=1\linewidth]{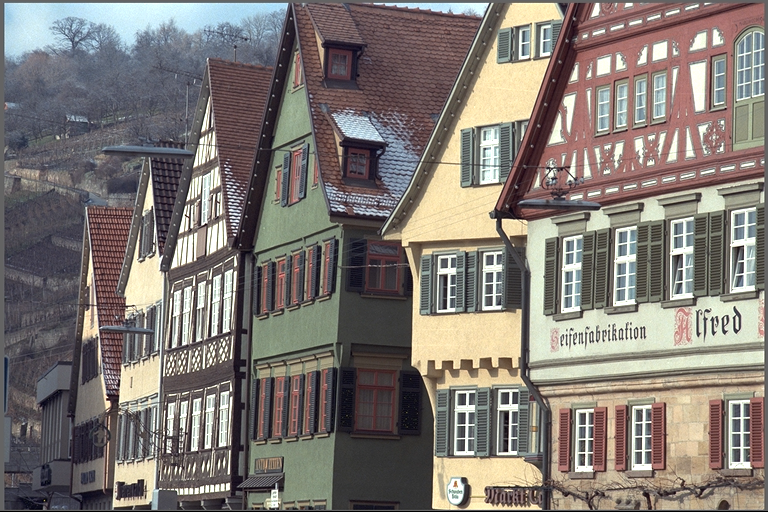}
			\end{minipage}%
		}%
		\quad
		\subfigure{
			\begin{minipage}[t]{0.29\linewidth}
				\centering
				\includegraphics[width=1\linewidth]{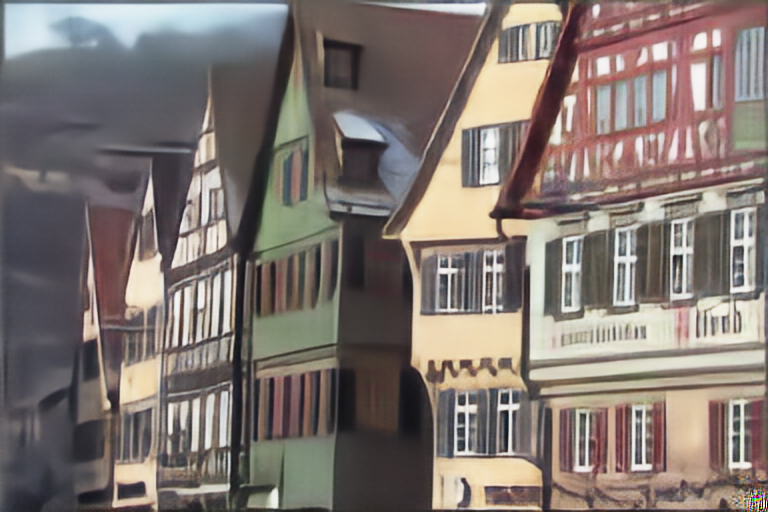}
			\end{minipage}%
		}%
		\quad
		\subfigure{
			\begin{minipage}[t]{0.29\linewidth}
				\centering
				\includegraphics[width=1\linewidth]{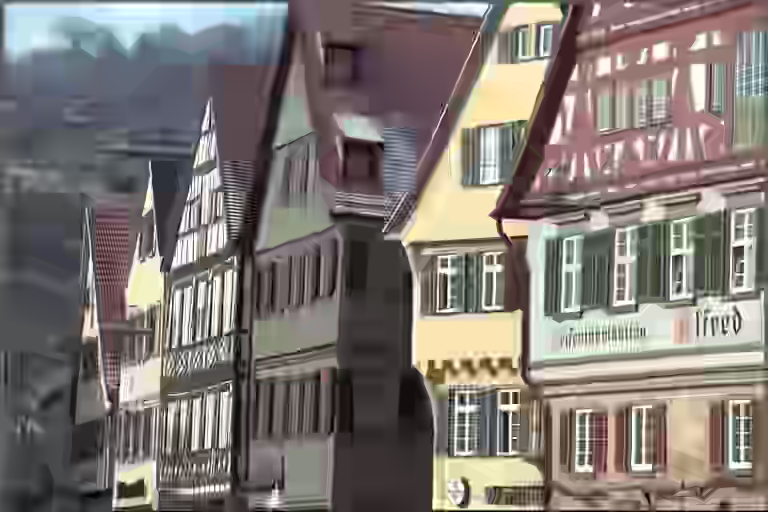}
			\end{minipage}%
		}%
		\centering
		\caption{{Visual comparison on Kodak image (left): example reconstructions by our MSAE framework (middle) and BPG (right). The respective bitrates, PSNR and SSIM are 0.070bpp, 19.78dB, 0.5386 vs 0.086bpp, 20.97dB, 0.6473.}}
	\end{figure}
	
	{\bf Performance Evaluation:}
	To evaluate the performance of our proposed MSAE based extreme image compression method, we compare our method with BPG and JPEG2000, as shown in Fig.~\ref{sfig:ADE20K}, where both objective PSNR, SSIM and subjective snapshots of two samples are illustrated.  For all the images we tested in datasets Cityscapes and ADE20K. We use basic arithmetic coding to generate actual bitstreams, and the bitrate is below 0.1 bpp. For quantitative evaluations, we compute the PSNR and SSIM between input $X$ and reconstruction $X^{'}$. But we have to mention that at such low bitrate, quantitative measurements such as PSNR or SSIM~\cite{wang2004image}  become meaningless as they penalize changes in local structure rather than the preservation of the global semantics.
	
	{It is clear that our method has demonstrated noticeable perceptual quality margin over traditional JPEG2000 and BPG, even with tiny loss in objective metrics like PSNR and SSIM. Similar conclusion can be found in Fig.\ref{sfig:ADE20K}.}
	This also coincides similar observations that learning based compression can usually provide better visual quality, but worse PSNR~\cite{balle2016end,Liu2018DeepCoder,agustsson2018generative}.
	
	

	\section{Concluding Remarks}
	\label{sec:conclusion}
	
	We have developed an extreme image compression framework via a multiscale autoencoder structure with embedded generative adversarial optimization for end-to-end training. Such multiscale authoencoder is fulfilled by downscaling the original image into various scales to capture the image statistics locally and globally.  Each decoded representation at lower resolution scale is utilized as the priors for the efficient compression at higher scale. In addition to traditional pixel-wise distortion measurements (e.g., PSNR, SSIM), we have introduced the adversarial loss for pleasant image reconstruction at a very low bitrate (i.e., usually below 0.05 bpp), to preserve the image structure and global semantics. Experimental studies have demonstrated that our method has provided subjective quality improvement over existing JPEG2000 and BPG on public datasets, but objective evaluation suffers (e.g., PSNR or SSIM). This calls for the interesting studies on quality metrics to accurately capture the visual significance in semantic domain for ultra low bitrate scenario.
	

	
	
	\bibliographystyle{IEEEbib}
	\bibliography{strings,refs}

\end{document}